\documentstyle[12pt,preprint,aps]{revtex}
 
\begin{document}
\input{epsf}
 
\preprint{UMHEP-434}
\draft
\title{How to trap a non-standard penguin?
Isospin Symmetry Violations in $B$-decays with 
Enhanced Chromomagnetic Dipole Operator.}
\author{
Alexey A. Petrov}
\address{
High Energy Theory Group \\
Department of Physics and Astronomy \\
University of Massachusetts \\
Amherst MA 01003}
\maketitle
\begin{abstract}
\noindent
A recently proposed Non-Standard Model solution to the 
problem of low semileptonic branching ratio $B_{SL}$ 
which suggests a large branching ratio for the decay 
$b \to sg$ is critically examined.
It is shown that the effects of the Enhanced Chromomagnetic Dipole Operator
might lead to significant violations of isospin symmetry in rare
radiative decays of $B$-mesons.  
\end{abstract}

\noindent
It has been recently proposed that effects of New Physics
associated with the gluonic dipole operator \cite{alex} 
provide an elegant and simple solution to the 
problem of low semileptonic branching ratio 
of $B$-mesons $B_{SL}$ simultaneously solving the 
problem of charm multiplicity $n_c$.
In these models, New Physics particles 
(e.g. techniscalars or supersymmetric particles)
running inside of the penguin diagram loop
significantly modify (increase) the Wilson coefficient $C_{11}$
in front of the gluonic dipole operator
$O_{11}=\frac{g_s}{16 \pi^2}
m_b \bar s \sigma^{\mu \nu} \frac{1}{2} 
(1-\gamma_5) t^a b G_{\mu \nu}^a$ in the effective Hamiltonian
\begin{equation}
H_{eff} = \frac{4 G_F}{\sqrt{2}} \Bigl[
V_{cb} V_{cs} ( C_1 (\mu) O_i(\mu)
C_2 (\mu) O_2(\mu) ) -
V_{tb} V^*_{ts} \sum 
C_i (\mu) O_i(\mu) \Bigr]
\end{equation}
increasing therefore the rate of the process $b \to s g$.
It is easy to see that increasing $\Gamma(b \to sg)$ by the effects
of New Physics we are simultaneously reducing  
$B_{SL}$ and $n_c$ as required by the experimental results:
\begin{eqnarray}
B_{SL} = \frac{\Gamma(B \to X l \nu)}{\Gamma_{tot}}, ~~~~~~
n_c = \frac{\Gamma(B \to X_c) + 2 \Gamma(B \to X_{c \bar c})}{\Gamma_{tot}},
\nonumber \\
\Gamma_{tot} = \sum_l \Gamma(B \to X l \nu) + 
\Gamma(B \to X_c )+ \Gamma(B \to X_{c \bar c}) + 
\Gamma(b \to sg,~ b \to s \gamma, ...).
\end{eqnarray}
 This, however,  leads to the modification of the already 
(and not yet) observed processes by Enhanced Chromo-Magnetic 
Dipole Operators (ECMDO).

\noindent
The direct effects of the ECMDO on the branching ratios
of the decays of the type $B^- \to \bar K \pi^-$ or 
$B^- \to X_s^- \phi$ (which are dominated by the gluonic 
penguin diagram) have been recently considered in 
\cite{antonio}.
Surprisingly enough, nowadays none of the {\em direct} measurements 
could show the difference in the predictions of the Standard Model 
and ECMDO models \cite{antonio}. 
The situation will certainly improve 
with new experiments and construction of the B-factory.
It is clear, however, that independent experimental tests are needed.

\noindent
One of the possible decay modes to serve as a test ground is 
the exclusive $B \to K^* \gamma$ decay (or relevant 
inclusive transition $B \to X_s \gamma$). The impact of 
ECMDO on this decay mode is
indirect, for instance, through the renormalization of the 
$b \to s \gamma$ vertex. The effect of it on the given branching ratio 
is relatively small, but might be larger on the portions of the
photon spectrum. This, however, requires some experimental efforts in 
pushing down the threshold in the observed $E_\gamma$.

\noindent
In this paper we consider another possibility.
Since hadrons are the asymptotic states of QCD there must
exist bound state corrections to the calculated decay rate
for $B \to K^* \gamma$ and $B \to X_s \gamma$.
These ``gluonic spectator effects'' in the exclusive 
$B\to K^* \gamma$ and inclusive $B \to X_s \gamma$
have been recently considered in \cite{milana} \cite{we}. It 
was shown that the charge (isospin) asymmetry 
\begin{equation} \label{ratio}
a = \frac{\Gamma_{B^- \to K^* \gamma} - \Gamma_{B^0 \to K^* \gamma}}
{\Gamma_{B^- \to K^* \gamma} + \Gamma_{B^0 \to K^* \gamma}}
\end{equation}
is fairly small in the Standard Model, typically of the order of 
a few percent. The primary sources 
of the isospin symmetry violations are long-distance effects (i.e. 
final state interactions or weak annihilation) \cite{dgp} 
and ``gluonic spectator effects''
(diagrams where photon is radiated off the spectator quark).
In the later case, the introduction of new 
gluonic penguin operator enhances these effects making the ratio 
(\ref{ratio}) to be a potentially sensitive test of the ECMDO models.
In what follows we consider isospin symmetry violations in the decay
$B \to K^* \gamma$ for different values of magnitude and phase of the 
$C_{11}^{ECMD}$ Wilson coefficient.

\noindent
We proceed with the estimation of the effect for the exclusive
transition $B \to K^* \gamma$ using Brodsky-Lepage (BL) perturbative QCD
formalism \cite{milana}. 
In this formalism the decay amplitude can be written as
\begin{eqnarray}
Amp = \int dx~ dy~ 
\phi^*_{K^*} (y) T_\mu (x,y) ~ {\xi^*}^\mu \phi_B (x), \nonumber \\
\phi^*_{K^*} = \sqrt{3} f_K y (1-y),~~~
\phi_B (x) = \frac{f_B}{2 \sqrt{3}} \delta (x - 1 + \epsilon_B), 
\nonumber \\
T_\mu (x,y) = \frac{1}{2} Tr \Bigl[
\rlap/{\epsilon}^* ( \rlap/{p}_{K^*} + m_{K^*}) t_\mu (x,y) 
\gamma_5 ( \rlap/{p}_B - m_B) \Bigr].
\end{eqnarray}
Here $x(y)$ is the momentum fraction carried by $b(s)$-quark inside of the
meson. The hard scattering amplitude $t_\mu (x,y)$ is calculated 
from the diagrams presented in the Fig.1 and the asymptotic 
expressions for the hadronic wave functions are used \footnote{
Although the use of asymptotic wave fuctions leads to the
underestimation of the branching ratios, hadronic 
uncertainties are cancelled to some extend in (\ref{ratio}).
Eventually one might study the effect using more model-independent
methods, e.g. QCD sum rules.}.
There are four classes of the spectator correction graphs involved
\cite{we}: \{1\} Photonic penguin diagrams. These give the leading 
contribution to the decay rate but do not involve gluonic dipole 
operators thus contributing to the asymmetry (\ref{ratio})
through the interference terms.
\{2\} ``Triangle'' and $W$-bremsstrahlung graphs. Their contributions 
are small compared to \{1\} and they do not modify the asymmetry.
We drop these contributions hereafter. 
\{3\} Bremsstrahlung diagrams involving photon emission
from external legs. These include spectator bremsstrahlung diagrams 
enhanced by large gluonic dipoles and thus largely responsible for the 
isospin symmetry breaking in the ECMDO models.  
\{4\} Weak annihilation graphs
that contribute to the decays of  $B^-$ but not
to $B^0$ mesons thus forming the
Standard Model ``background'' to the asymmetry.
 
\noindent
We calculate the branching ratios according to the power
counting of \cite{milana} modified for the leading non-zero $K^*$ mass effects.
In particular, since the main focus here is the estimation of the isospin 
asymmetry, terms up to the order of $m_{K^*}/(m_B ~\epsilon_B)$ 
(scale of the weak annihilation contribution) \cite{milana},  
$\epsilon_B=\bar \Lambda / m_B \sim 0.065-0.1$ must be included.
The power counting is governed by the peak approximation to 
$B$-meson distribution function
and by the expansion of the pQCD amplitude in powers of $1/m_B$: 
terms $1/\epsilon_B$  
scale like $m_B^1$ and $m_{K^*}/(m_B \epsilon_B)$ scale like
$m_B^0$, so we must keep all of the terms up to the order $m_B^0$.
\footnote{In principle, any term proportional to 
$1/\epsilon_B$ gives $m_B/ \bar \Lambda$ proportional to the 
non-perturbative parameter $\bar \Lambda^{-1}$, thus
bringing non-perturbative uncertainty to the 
pQCD calculation. One can fix this 
uncertainty by fixing the value of $\bar \Lambda$ from other $B$-decays.} 
In this calculation we systematically neglect terms $\sim
m_{K^*}^2$ with respect to the terms $\sim m_{K^*}$ in the amplitude. 
Gauge invariance implies that the decay rate can be written as
\begin{equation}
\Gamma ( B(p_B) \to K^* (p_{K^*}) ~\gamma(q)) = 
\frac{1}{16 \pi} \frac{m_B^2 - m_{K^*}^2}{m_B^3} 
( |a|^2 [3 - \frac{y}{x^2} (2 x  - 1)]
+ 2 |c|^2 x^2 )
\end{equation}
with $x=p_B \cdot q/m_B^2$, $y=1-m_B^2 (1-x)^2/m_{K^*}^2$, and 
$a(c)=\sum_i a_i(c_i)$, and
\begin{equation}
t_{i~ \mu \nu} = a_i \bigl[ g_{\mu \nu} - 
\frac{1}{x m_B^2} p_\mu p_\nu \bigr]
- \frac{i c_i}{m_B^2}~ 
\epsilon_{\mu \nu \alpha \beta}~ p_\alpha q_\beta
\end{equation} 
The dominant contribution comes from the diagrams of the class \{1\}.
\begin{eqnarray}
t_1 = \frac{i g_s^2 C}{2 Q^2 k_1^2}~
{\cal T}r ~  \rlap/{\epsilon^*} ( \rlap/{p}_{K^*} + m_{K^*})
\gamma_\alpha \rlap/{k}_1 V_{10} \cdot \xi^* 
\gamma_5 (\rlap/{p}_B - m_B ) \gamma_\alpha , 
\nonumber \\
t_2 = \frac{i g_s^2 C}{2 Q^2 
(k_2^2 - m_B^2(1-\epsilon_B)^2)}~~~~~~~~~~~~~~~~~~~~~~~~~ 
\nonumber \\
{\cal T}r ~  \rlap/{\epsilon^*} ( \rlap/{p}_{K^*} + m_{K^*})
V_{10} \cdot \xi^* [\rlap/{k}_2 + m_B(1-\epsilon_B)]
\gamma_\alpha \gamma_5 (\rlap/{p}_B - m_B ) \gamma_\alpha ,
\end{eqnarray}
where $C=\frac{16 q_{up}}{3 \sqrt{2}} G_F V_{tb} V_{ts}^* 
C_{10}^{eff}$, $C_{10}^{eff} = -0.32$, 
$V_{10 \alpha}= m_b ~i \sigma_{\alpha \beta} (1 - \gamma_5) q_\beta$, and 
$Q_\alpha$ is the momentum of a gluon. Also,
\begin{eqnarray}
k_1^2= -m_B^2 ~x (1-x),
\nonumber \\
k_2^2-m_b^2 = - m_B^2 (1 - y - 2 \epsilon_B),
\nonumber \\
Q^2 = -m_B^2 (x-y) (1-x).
\end{eqnarray}
Note that $m_b = m_B (1- \epsilon_B)$. Calculating the traces we arrive at
\begin{eqnarray}
a_1= -\frac{1}{4} \frac{\alpha_s}{2 \pi} C f_B f_{K^*} m_B 
\frac{m_{K^*}}{\epsilon_B m_B} , 
\nonumber \\
c_1= -\frac{1}{2} \frac{\alpha_s}{2 \pi} C f_B f_{K^*} m_B 
\frac{m_{K^*}}{\epsilon_B m_B} ,
\nonumber \\
a_2=-\frac{\alpha_s}{2 \pi} C f_B f_{K^*} 
\frac{m_B}{4 \epsilon_B} 
\Bigl\{ \epsilon_B +
\frac{m_{K^*}}{m_B} -
4 \epsilon_B \log \frac{1-2 \epsilon_B}{2 \epsilon_B}
+ 2 \epsilon_B \log \frac{1-\epsilon_B}{\epsilon_B} - 2 i \pi 
\Bigl \} ,
\nonumber \\
c_2=-\frac{\alpha_s}{2 \pi} C f_B f_{K^*} 
\frac{m_B}{2 \epsilon_B} 
\Bigl\{ - \epsilon_B +
 \frac{m_{K^*}}{m_B} +
4 \epsilon_B \log \frac{1-2 \epsilon_B}{2 \epsilon_B}
- 2 \epsilon_B \log \frac{1-\epsilon_B}{\epsilon_B} + 2 i \pi 
\Bigl \} .
\end{eqnarray}
The diagram 1 of the class \{1\} was dropped in \cite{milana} in
the approximation $m_{K^*}=0$. 
The major contribution to the asymmetry (\ref{ratio}) comes from the 
interference of the diagrams of class \{3\} 
(especially those involving 
spectator bremsstrahlung) and class \{1\} where the
former includes the  vertex for $bsg$ $V^{11}_\alpha$:
\begin{equation}
V^{11}_\alpha = F_1 (Q^2 g_{\alpha \beta} - Q_\alpha Q_\beta)
\gamma_\beta (1 + \gamma_5) + F_2 m_b i \sigma_{\alpha \beta} Q_\beta 
(1 - \gamma_5)
\end{equation}
with $F_i$ being the QCD-corrected Inami-Lim functions,
$F_2 = C_{11}$. In the Standard 
Model $C_{11}^{SM}=-0.159$, $|C_{11}^{ECMD}| \approx 7 |C_{11}^{SM}|$.
The denominators are 
\begin{eqnarray}
k_5^2=(1-y)~m_B^2~ \Bigl\{
1-y(1-y) \frac{m_{K^*}^2}{m_B^2} 
\Bigr\} \to (1-y)~m_B^2 ,
\nonumber \\
k_6^2 = (1-x)~m_B^2 ~ \Bigl\{
-x + \frac{m_{K^*}^2}{m_B^2} \Bigr\}
\to -x(1-x)~m_B^2 ,
\nonumber \\
Q^2=m_B^2 ~ (x-y) \Bigl\{
x - \frac{m_{K^*}^2}{m_B^2} \Bigr\} \to
m_B^2~x (x-y) .
\end{eqnarray}
The formfactors for the spectator bremsstrahlung diagrams
read
\begin{eqnarray}
a_5=-q_{u,d} C' m_B 
f_B f_{K^*} \frac{\alpha_s}{2 \pi}
\Bigl \{ 
- \frac{3}{4} F_1 + \Bigl[ -\frac{3}{4} 
+ \log \frac{1-\epsilon_B}{\epsilon_B} + i \pi \Bigr ] F_2
\Bigr \} ,
\nonumber \\
c_5=-q_{u,d} C' m_B f_B f_{K^*} 
\frac{\alpha_s}{2 \pi} 
\Bigl \{ - \frac{3}{2} F_1 + \Bigl[ \frac{3}{2} 
-2 \log \frac{1-\epsilon_B}{\epsilon_B} - 2 i \pi \Bigr ] F_2 \Bigr \} ,
\nonumber \\
a_6=-\frac{q_{u,d} C' m_B}{1-2 \epsilon_B} 
\frac{\alpha_s}{2 \pi} f_B f_{K^*}
\frac{1}{2 \epsilon_B}
\Bigl \{ F_1 \Bigl [ \frac{1}{2} \epsilon_B + \frac{1}{6} 
\frac{m_{K^*}}{m_B} \Bigr ]
+ F_2 \Bigl[ \frac{1}{2} \epsilon_B + \frac{1}{6} 
\frac{m_{K^*}}{m_B} \Bigr ]
\Bigr \} ,
\nonumber \\
c_6=-\frac{q_{u,d} C' m_B}{1-2 \epsilon_B} 
\frac{\alpha_s}{2 \pi} f_B f_{K^*}
\frac{1}{\epsilon_B}
\Bigl \{ F_1 \Bigl [ -\frac{1}{2} \epsilon_B - \frac{1}{6} 
\frac{m_{K^*}}{m_B} \Bigr ]
- F_2 \Bigl[ \frac{1}{2} \epsilon_B + \frac{1}{6} 
\frac{m_{K^*}}{m_B} \Bigr ]
\Bigr \} .
\end{eqnarray}
with  $C'=\frac{4}{3 \sqrt{2}} G_F V_{tb} V_{ts}^*$.
In the Standard Model, the contribution of the 
operator 
\begin{equation}
O_{11}=\frac{g_s}{16 \pi^2}
m_b \bar s \sigma^{\mu \nu} \frac{1}{2} 
(1-\gamma_5) t^a b G_{\mu \nu}^a
\end{equation}
is suppressed, mainly because of the numerical smallness
of the $C_{11}$ coefficient in comparison with 
the photonic dipole coefficient $C_{10}$:
in the SM $C_{10}/C_{11} \approx 2$. In addition,
bremsstrahlung diagrams are suppressed dynamically.
In the ECMDO, the $C_{11}$ is approximately one order 
of magnitude higher, therefore compensating to some extend
the dynamical suppression. 

\noindent
Clearly, the relative phase
between the ECMDO and SM $C_{11}$'s is not fixed, thus
leaving some freedom, and, in principle, it can be chosen in a 
way that it does not affect $B \to K^* \gamma$ SM predictions.
This calculation, however, indicate that the isospin asymmetry
is large on the large portion of the available parameter space therefore
providing a good constraint on its value. The results of 
calculations are presented in the Fig.2. 

\noindent
Unfortunately, the isospin asymmetry has not yet been a target for the
experimental investigation. 
Recent CLEO data \cite{cleo}, for instance, 
gives $Br_{B^- \to K^* \gamma} = 3.8 ^{+2.0}_{-1.7} \pm 5.0$ and
$Br_{B^0 \to K^* \gamma}=4.4 \pm 1.0 \pm 0.6$
providing only 
a rough estimate of the isospin symmetry violations in the
range $0-50\%$ which is clearly unsatisfactory for 
singling out the ECMDO contribution. 
Hopefully, combined
results from the isospin violation in $B \to K^* \gamma$ and
direct measurements would put strong constraints on the values of
contributions from ECMDO models.

\begin{figure}
\centerline{
\epsfysize 3.5in
\epsfxsize 4in
\epsfbox{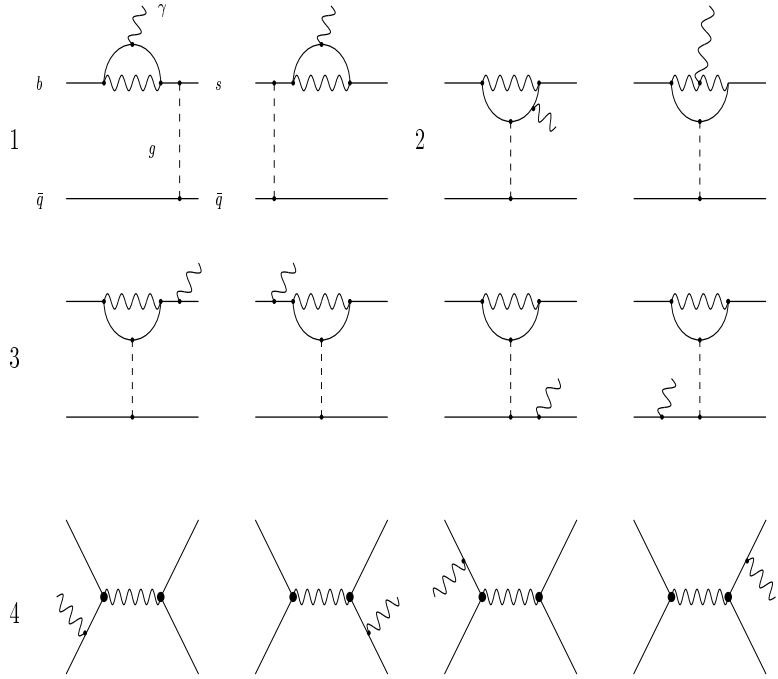}}
\caption{Diagrams for the gluonic spectator corrections.}
\end{figure}

\begin{figure}
\centerline{
\epsfbox{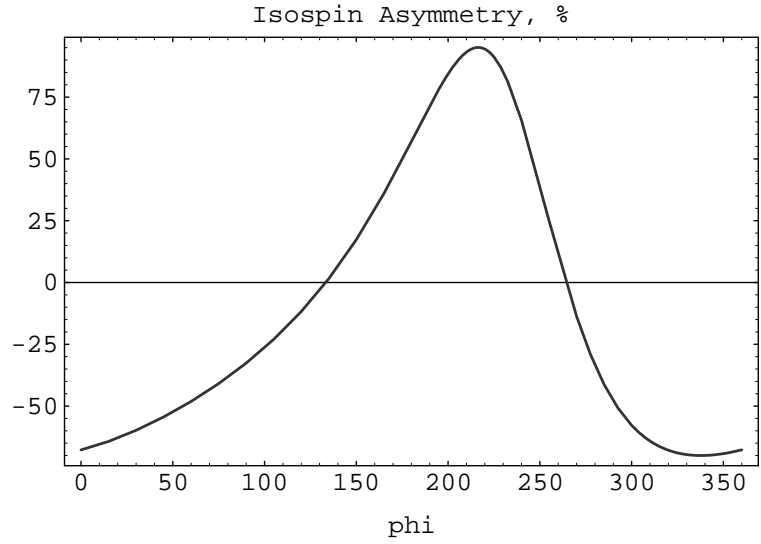}}
\caption{Isospin violation asymmetry for different values of phase between
SM and ECMDO coefficients.}
\end{figure}


\begin{references}

\bibitem{alex}{A. L. Kagan, Phys. Rev. {\bf D51} (1995) 6194;
M. Ciuchini, E. Gabrielli, 
G.F. Giudice, Phys. Lett. {\bf B388} (1996) 353;see also
B. Grazadkowski, W.-S. Hou, Phys. Lett. {\bf B272} (1991) 383.}

\bibitem{antonio}{A. L. Kagan, A. F. Perez, in preparation; 
A. F. Perez, Talk delivered at Phenomenology-96 Symposium, Wisconsin,
unpublished; 
A. L. Kagan, Talk delivered at DPF-96 Meeting, Minnesota, to be published
in the proceedings.}

\bibitem{milana}{C.E. Carlson, J. Milana, Phys. Rev. {\bf D51} (1995) 4950.}

\bibitem{we}{J.F. Donoghue, A.A. Petrov, Phys. Rev {\bf D53} (1996) 3664.}

\bibitem{dgp}{J.F. Donoghue, E. Golowich, A.A. Petrov, 
Preprint UMHEP-433, hep-ph/9609530, to be published in Phys. Rev. D;
A. Ali, V.M. Braun, Phys. Lett. {\bf B359} (1995); H.-Y. Cheng, 
Phys. Rev. {\bf D51} (1995).}

\bibitem{cleo}{R. Kass, ``Recent results from CLEO'', 
Talk delivered at 1996 SLAC Summer Institute, to be published
in the proceedings.}
\end{references}
\end{document}